\documentclass{article}

\usepackage{arxiv}

\usepackage[utf8]{inputenc} 
\usepackage[T1]{fontenc}    
\usepackage{hyperref}       
\usepackage{url}            
\usepackage{booktabs}       

\usepackage{threeparttable}
\usepackage{array}
\usepackage{makecell}

\usepackage{amsfonts}       
\usepackage{nicefrac}       
\usepackage{microtype}      
\usepackage{lipsum}		
\usepackage{graphicx}
\usepackage{natbib}
\usepackage{doi}

\usepackage{amsmath}
\usepackage{amsfonts}
\usepackage{amssymb}
\usepackage{amsbsy}
\usepackage{amsthm}
\usepackage[nolist]{acronym}
\usepackage{eurosym}
\usepackage[table]{xcolor}
\usepackage{cleveref}
\usepackage{subcaption}

\title{Comparison of Energy System Optimization Software and Evaluation of Selected Frameworks}

\date{} 					

\author{ {\hspace{1mm}Pedro~Caixeta} \\
	Institute for Automation and Applied Informatics\\
	Karlsruhe Institute of Technology\\
	76131 Karlsruhe, Germany \\
	\texttt{ucwdb@student.kit.edu} \\
        \And
	{\hspace{1mm}David~Gawron} \\
	Institute for Automation and Applied Informatics\\
	Karlsruhe Institute of Technology\\
	76131 Karlsruhe, Germany \\
	\texttt{undhh@student.kit.edu} \\
        \And
        \href{https://orcid.org/0000-0002-1463-7606}{\includegraphics[scale=0.06]{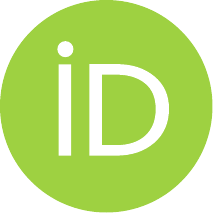}\hspace{1mm}Hüseyin K.~Çakmak} \\
	Institute for Automation and Applied Informatics\\
	Karlsruhe Institute of Technology\\
	76131 Karlsruhe, Germany \\
	\texttt{hueseyin.cakmak@kit.edu} \\
        \And
        \href{https://orcid.org/0009-0005-0067-8019}{\includegraphics[scale=0.06]{orcid.pdf}\hspace{1mm}Haozhen~Cheng} \\
	Institute for Automation and Applied Informatics\\
	Karlsruhe Institute of Technology\\
	76131 Karlsruhe, Germany \\
	\texttt{haozhen.cheng@kit.edu} \\
}

\renewcommand{\shorttitle}{Comparison of Energy System Optimization Software}

\hypersetup{
pdftitle={Comparison of Energy System Optimization Software and Evaluation of Selected Frameworks},
pdfsubject={eess.SY},
pdfauthor={Pedro~Caixeta, David~Gawron, Hüseyin K.~Çakmak, Haozhen~Cheng},
pdfkeywords={},
}

\begin{document}

\begin{acronym}
    \acro{KIT}{Karlsruhe Institute of Technology}
    \acro{FZJ}{Forschungszentrum Jülich}
    \acro{DH}{District Heating}
    \acro{TABULA}{Typology Approach for Building Stock Energy Assessment}
    \acro{DHW}{Domestic Hot Water}
    \acro{HIU}{Heat Interface Unit}
    \acro{HVAC}{Heating, Ventilation and Air Conditioning}
    \acro{FMI}{Functional Mock-up Interface}
    \acro{FMU}{Functional Mock-up Unit}
    \acro{LPG}{LoadProfileGenerator}
    \acro{PV}{Photovoltaic}
    \acro{GUI}{Graphical User Interface}
\end{acronym}

\maketitle

\begin{abstract}
    Optimizing energy systems is a crucial step toward achieving a carbon-neutral future, with software tools playing a major role in the process. However, selecting the most suitable tool for specific optimization challenges can be complex, given the diverse objectives and requirements of various energy systems. In this study, we aim to address this issue by evaluating five preselected software tools—REMix, MTRESS, COMANDO, OEMOF, and HOMER PRO—to identify the scenarios for which they are most suitable. To achieve this, we conducted an extensive review of literature, documentation, tutorials, and example models, and developed a set of comparison criteria that were subsequently used to evaluate these tools. Our analysis suggests that REMix is particularly effective for scenarios where investment path optimization is required. COMANDO excels in systems with atypical components. OEMOF is the best-suited open-source tool for standard optimization problems. HOMER PRO is recommended for users seeking rapid synthesis optimization, especially those with limited programming experience. Due to difficulties in obtaining sufficient information on MTRESS, we were unable to complete the analysis for this tool.
\end{abstract}


\section{Introduction}
\label{Motivation}

The motivation for this work stems from a persistent trilemma that has accompanied society for years and will continue to do so in the foreseeable future: How can we achieve economic growth without depleting resources (such as energy) and without causing environmental degradation \cite{FRANGOPOULOS20181011}? This challenge is particularly evident in the context of current and future energy systems, which must be optimized to minimize resource depletion and environmental impact. In this context, optimization can be understood as the process of finding the best way to achieve a particular objective (or multiple objectives) under given constraints \cite{OptimizationEngeneeringBook}. As for an energy system, it can be defined as a system that transforms energy between different forms and/or transfers energy between different places \cite{FRANGOPOULOS20181011}.

The optimization of an energy system can be categorized into three levels: Synthesis, Design and Operation.
\begin{itemize}
  \item Synthesis optimization describes the optimal composition of the system, including adding and removing components and interconnections. 
  \item Design optimization focuses on optimizing the technical characteristics (specifications) of the components and the properties of the substances that flow in and out of each component.
  \item Operation optimization concerns identifying the optimal operation point under specified conditions in a given system with finished synthesis and operation optimization. \cite{FRANGOPOULOS20181011}.
\end{itemize}

Thus, to optimize a system means to define its synthesis, the design characteristics of its components and the operating point that leads to an overall optimum \cite{articleFrangopoulos}.

In real-world scenarios, energy system optimization often involves multiple levels and objectives, making the task highly complex. As a result, the use of software tools becomes essential \cite{CONNOLLY20101059}. However, identifying the ideal software tool for a specific optimization problem can be challenging and time-consuming, as the selection process heavily depends on the specific objectives of the problem to be solved.

\section{Literature Review and Contribution}
This section provides an explanation of how this study connects with existing literature in Subsection \ref{Related Work}, followed by a clear definition of our research objective in Subsection \ref{Purpose of this Study}.

\subsection{Related Work} \label{Related Work} 
Primarily, four distinct categories of literature have been used in this project. In this section, their relevance to the project topic is explained. These four literature types are:

\begin{enumerate}
  \item Literature presenting the software packages that are the study topic of this paper;
  \item Literature describing various works, projects, research, or studies, that utilized at least one of the five tools examined;
  \item Literature comparing software tools that are used in the energy field; and
  \item Literature explaining the optimization problems of energy systems.
\end{enumerate}

\subsubsection{Introduction of the software packages and application examples}
The first two categories of literature are essential because this project focuses on researching software tools. Thus, literature that presents these software tools and their real-world applications is particularly valuable as a source of information. For example the book by \cite{oemofBook} compiles key documentation from the OEMOF-Community, making it accessible to the user. One of the examples of OEMOF application described in this book is the Study "Electrification of Agricultural Machinery and Use of Photovoltaics", where OEMOF has been used to calculate investment costs.

Similarly, the paper by \cite{GILS2017173} introduces REMix, its approach and its functions. The study applies REMix to a case study to research the correlation between variable renewable energy penetration, balancing power demand and supply costs of a set of 19 experimental energy systems scenarios.

MTRESS is best presented in the document of \cite{schönfeldt2022mtress30modell}. 
An example of a use case is the topic of the book of \cite{PrefaceDeGruyter}. MTRESS was used to simulate and optimize supply concepts considering energy costs, CO2 emissions and own consumption using the example of a neighborhood.

A reference for COMANDO is the paper of \cite{LANGIU2021107366}. It introduces COMANDO, its approach regarding modeling, problem formulation and problem-solving. It also demonstrates key features of COMANDO in four case studies. One of these case studies involves the design of a new industrial energy system.

As for HOMER PRO, an introduction to the software can be found directly in its \ac{GUI} and some case studies are approached in the paper of \cite{CONNOLLY20101059}. There it is mentioned that HOMER PRO has been used to determine the feasibility of a stand-alone wind-diesel hybrid in Saudi Arabia, to assess the wind energy potential at individual locations in Ethiopia and to simulate a stand-alone system with hydrogen in Canada.

\subsubsection{Comparing software tools in the energy field}
Comparative literature, the third category, provides valuable methodologies and approaches that can be used as a basis for the comparative analysis conducted in this project. For instance, the paper of \cite{CONNOLLY20101059} reviews different computer tools, including HOMER PRO, for analyzing renewable energy integration. The methodology involved communication with software developers through surveys. A literature work that compares some of the 5 software tools could also be useful in this project, but none could be found.

\subsubsection{Optimization of energy systems}
Lastly, the fourth category of literature provides essential information on optimizing energy systems, forming a solid foundation for this project's focus on energy system optimization. The paper of \citeauthor{FRANGOPOULOS20181011} \cite{FRANGOPOULOS20181011} was particularly relevant. It includes definitions of optimization and energy systems, as well as discussions on trends and challenges in energy system optimization.

\subsection{Purpose of this Study} \label{Purpose of this Study}

Despite the extensive literature and numerous papers available, no existing work provides a comparative analysis of the software tools REMix \cite{REMixsoftware}, MTRESS \cite{MtressHelmholtz}, COMANDO \cite{ComandoHelmholtz}, OEMOF \cite{OemofWebsite}, and HOMER PRO \cite{HomerPROWebsite}. This gap in the literature is the motivation behind this study and the resulting paper.

The primary objective of this work is to compare five software tools to facilitate their selection for an energy system optimization task. Specifically, for each tool, we aim to answer the question: For which scenario is this tool the most suitable? This means that we will evaluate and compare the preselected software tools to determine the application context in which each of these tools is best suited.

The approach used to address this question is detailed in Section \ref{Method}. The results obtained are analyzed in Section \ref{Evaluation} and presented in Section \ref{Presentation of the Results from the Comparison of the Existing Models}. In Section \ref{Discussion} a discussion of our approach and the obtained results takes place. We conclude our work and present future outlooks in Section \ref{Conclusion}.

\section{Method} \label{Method}
This study is conducted in two distinct phases. The approach of the first phase, which includes an initial comparison (that serves as an introduction to the researched software packages) is presented in Subsection \ref{Literature research section}. The approach of phase two, which results in the final comparison, is the subject of Subsection \ref{Comparison of the Existing Models}.

\subsection{Introduction to the Researched software Tools and First Comparison} \label{Literature research section} 
In the first phase, a comprehensive literature review was conducted to gather information about the five software tools to introduce those in a structured manner. This makes the first comparison between the software tools possible and highlights some differences between them. 
Some of the gathered information from our literature research is displayed in Table \ref{introduction of sw}. This table contains for each of the five software tools:

\begin{itemize}
  \item a short description of their goal (according to their developers),
  \item the number of official tutorials (also referred to as "example models" further in this paper) available to the users,
  \item their license costs, and
  \item their interface.
\end{itemize}

During this first analysis, we encountered some difficulties in finding information on MTRESS leading to its exclusion from further analysis in the second phase. More concretely, these difficulties are:
\begin{itemize}
  \item the release version 2.2.0 only offers one example and no documentation, and
  \item the newer developer version v3.0.0a2 offers one introduction paper for MTRESS and four examples, that failed to run with the expected versions of the dependencies. 
\end{itemize}

\begin{table*}
\centering
\renewcommand{\arraystretch}{1.5}
\begin{tabular}{@{} p{2cm} p{7cm} c p{3cm} c @{}}
\toprule
{Software} & \multicolumn{1}{c}{Goal} & \multicolumn{1}{>{\centering\arraybackslash}m{15mm}}{Number of Tutorials} & \multicolumn{1}{c}{License} & \multicolumn{1}{c}{Interface} \\

\midrule

REMix &  Set up linear optimization models in the field of energy systems modeling \cite{REMixsoftware}
 & 5  & Free, but solver costs 3500\$ & Python \\

MTRESS &   Facilitate the creation of models of residential energy supply systems, that can be linearly optimized  \cite{mtresssoftware}
& 1 & Free  & Python \\

COMANDO &   Object-oriented modeling of energy systems, creating and solving problems related to the model \cite{LANGIU2021107366}
& 5  & Free  & Python \\

OEMOF &   Provide a toolbox for the modeling of energy systems, create a function related to it that is to be minimized \cite{oemofBook}
& 32   & Free  & Python \\

HOMER PRO &   Optimize microgrid design by evaluating different model variants  \cite{HomerPROWebsite}
& 24
 & Licensed, with free trial  & Own interface\\

\bottomrule

\end{tabular}
\caption{Overview from the results of the literature research}
\label{introduction of sw}

\end{table*}

\subsection{Final Comparison} \label{Comparison of the Existing Models} 

The approach for the second phase of this work is divided into three main steps, as outlined in Subsection \ref{evaluation of existing models}, Subsection \ref{Development of comparison criteria} and Subsection \ref{Final Comparison}.

\subsubsection{Evaluation of Existing Models}\label{evaluation of existing models}
The first step of the approach comprises a broad analysis of all official model examples from each software tool (in total 62). These models (except for those that didn't result in new knowledge gain, for instance, due to their high similarity to an already evaluated model) were thoroughly investigated and examined. Key aspects evaluated during this step include:
\begin{itemize}
  \item the components that constitute the model,
  \item the connection between these components, and
  \item the optimization problem for which the model was developed and their objectives.
\end{itemize}

\subsubsection{Development of Comparison Criteria}\label{Development of comparison criteria}
\begin{figure}[ht]

\centering
\includegraphics[width=0.5\textwidth]{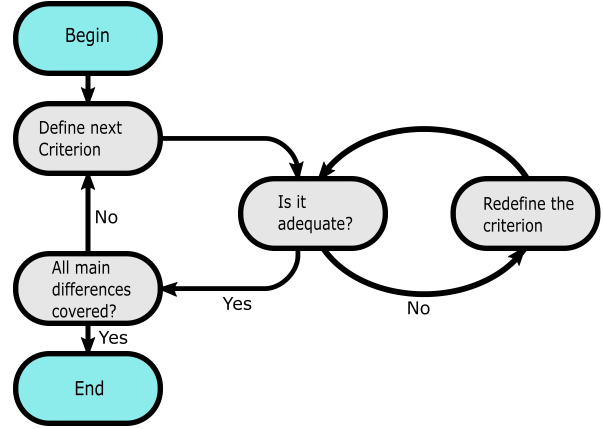}
\caption{Schematic representation of our work process to develop comparison criteria.}
\label{Workflow Figure}
\end{figure}

In the second step, the information gathered from the analysis of the software packages was used to develop appropriate comparison criteria. This process was iterative and is displayed in Figure \ref{Workflow Figure}. We employed a trial-and-error approach: criteria were initially defined based on differences between the software tools identified in the previous analysis, then their adequacy was assessed and subsequently refined until they met the necessary standards. For a comparison criterion to be deemed adequate, it must:
\begin{itemize}
  \item highlight clear and concrete differences between the software tools
  \item and be relevant for an individual seeking the most suitable software tool for optimizing their energy system.
\end{itemize}

Criteria that did not reveal significant differences nor offer relevant insights for users were discarded. Fine criteria or criteria that heavily depend on each other are aggregated. This iterative process was repeated until the main differences between the software tools, that we could investigate considering the limited workload available for this project, were adequately captured by the comparison criteria. The finalized comparison criteria are presented in Subsection \ref{Introduction to the comparison criteria}.

\subsubsection{Final Comparison}\label{Final Comparison}
In the third and final step, the software tools OEMOF, COMANDO, REMix, and HOMER PRO were evaluated using the comparison criteria developed in the previous step (see Subsection \ref{Development of comparison criteria}), thereby concluding the comparison between them. The results of this evaluation are presented in Section \ref{Presentation of the Results from the Comparison of the Existing Models}.

\section{Evaluation Criteria} \label{Evaluation}
In this Section, the procedure for the analysis of the results is introduced in Subsection \ref{Evaluation Criteria}. The outcomes from the analysis are then presented in the Subsection \ref{Analysis}.

\subsection{Methodology for Analysis}\label{Evaluation Criteria}
In this Section, we explain how the results from 'phase 2' of this work were analyzed to identify the scenarios in which each researched software tool excels. To achieve this, we first selected one software tool, and, using the results presented in Section \ref{Presentation of the Results from the Comparison of the Existing Models}, we searched for the characteristics that distinguish it the most from the other tools. The optimal use case for this software is then derived from these criteria, as it should involve an optimization problem where these characteristics play a significant role. This process was repeated for the remaining three software tools, and the results are summarized in Subsection \ref{Analysis}.

\begin{table*}
\centering
\renewcommand{\arraystretch}{1.5}
\begin{tabular}{@{} p{2cm} p{6cm} p{6cm} @{}}
\toprule
{Software} & \multicolumn{1}{c}{Key Characteristic} & \multicolumn{1}{c}{Best Suited For}\\

\midrule

REMix &  Optimize investment path, long-term economic optimization  & Optimization of long-term economic problems of large energy systems.   \\

COMANDO &  High level of detail, widest range of components & Scenarios requiring such a level of detail and modeling of unusual components.   \\

OEMOF &   Open Source (including solvers), simple and user-friendly Python programming & Standard optimization problems with energy systems. \\

HOMER PRO &   Built-in comparison of different variants of an energy system & Users with limited programming experience and who need to draw conclusions quickly. \\

\bottomrule

\end{tabular}
\caption{Overview of the findings from the analysis of the results}
\label{table analysis}

\end{table*}

\subsection{Findings from the Analysis}\label{Analysis}
This subsection presents the analysis of the results based on the procedure outlined in Subsection \ref{Evaluation Criteria}. Each of the following paragraphs comprises the analysis of one of the four software tools under study. An overview of the findings is displayed in Table \ref{table analysis}.

The analysis begins with REMix, which is distinguished by its capability to optimize the investment path, which offers a significant advantage in optimization problems that involve flexible investment paths—commonly encountered in the long-term economic optimization of large energy systems.

Next, COMANDO is evaluated. Its greatest advantage is its level of detail, which allows the widest range of components to be modeled. A notable example identified during the analysis of existing models involved researching the heat flow within one building, considering components such as the heat pump, air mass, and wall mass. Thus, this software tool excels in scenarios requiring such a level of detail and modeling of unusual components.

The key characteristic of OEMOF is that, among the software tools available for full use at no cost (OEMOF and COMANDO), it offers the highest level of user-friendliness and simplicity in model creation. Consequently, this software package is particularly well-suited for standard optimization problems, as it minimizes both financial costs and the operational effort required.

Finally, HOMER PRO excels in comparing different variants of an energy system, making it the preferred choice for projects where the energy system can undergo significant changes—provided these variants can be modeled within the software's scope (see Subsection \ref{modeling scope} for detailed information). Additionally, due to its graphical user interface and ease of use, HOMER PRO is particularly suitable for users with limited programming experience in Python who need to quickly conclude their projects.

\section{Comparison Results} \label{Presentation of the Results from the Comparison of the Existing Models}
In this section, we first briefly introduce the final criteria for the comparison of the software tools. Then, each criterion is further explained and all of the four software packages are evaluated according to it.

\subsection{Introduction to the Comparison Criteria}\label{Introduction to the comparison criteria}

\begin{figure}[ht]

\centering
\includegraphics[width=0.5\textwidth]{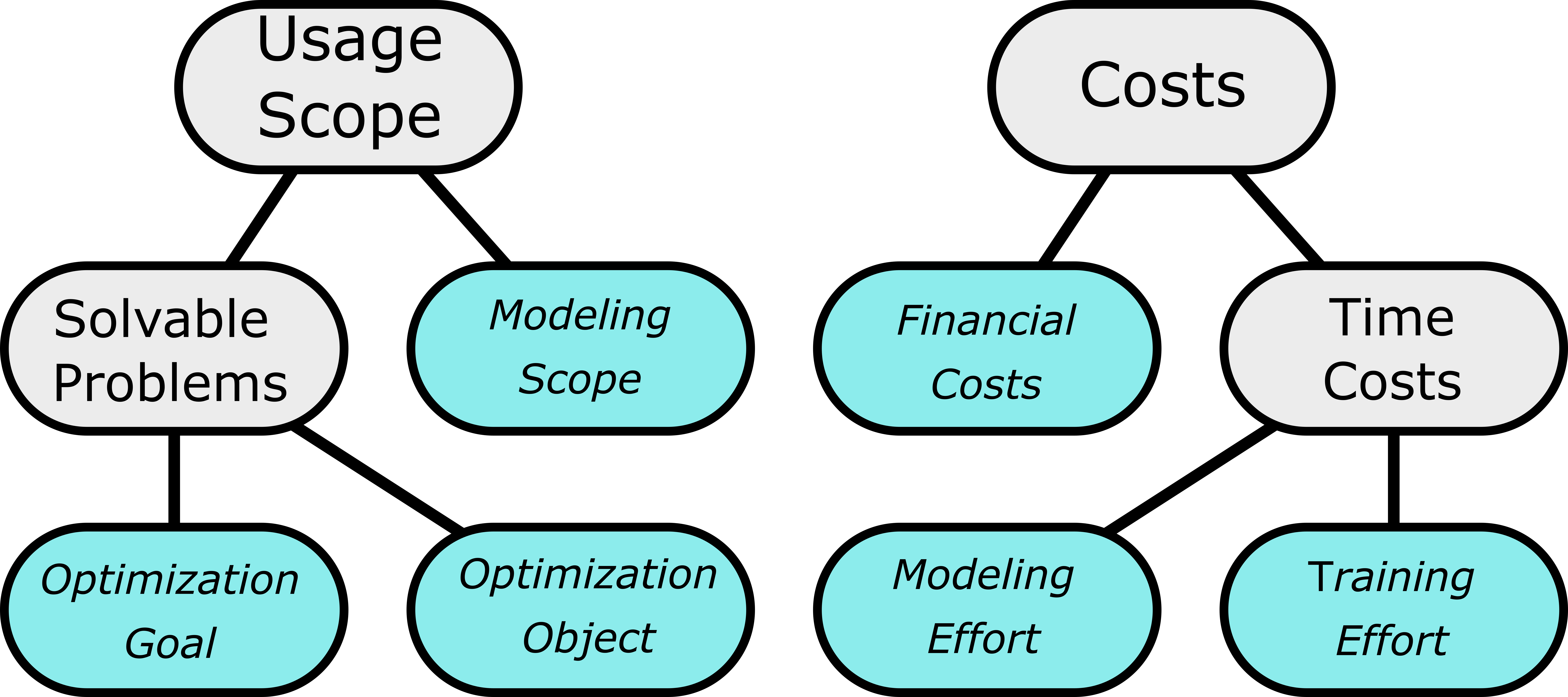}
\caption{Cathegorization of the comparison criteria. The blue boxes mark the 6 final criteria.}
\label{Criteria Division Figure}
\end{figure}

The differences between the four software packages studied in Phase 2 of this project are categorized into two main groups: effort and usage scope. The effort topic is further divided until it reaches three final criteria: financial costs, training effort and modeling effort. Similarly, the usage scope topic is subdivided into three final criteria: optimization goal, optimization object and modeling scope. Figure \ref{Criteria Division Figure} shows this division. In total, we have six final criteria, that are separately regarded in the following subsections.

\subsection{Financial Costs}\label{Financial Costs}
The costs associated with using the software tools depend mostly on the price of their license. In this regard, OEMOF and COMMANDO are the only tools that can be used freely without a license. To fully use them, the user needs access to a solver which, unlike REMix, could be open source. REMix depends exclusively on GAMS, which requires a license to function.
Although GAMS offers a demo and an academic license, that grants free access to part of this solver's functionalities, it is too limited to run REMix's tutorials. Consequently, the user is obliged to acquire a license for GAMS to run a REMix model. The basic GAMS module costs 3500 US\$ but the price can be reduced by 80\% for academic purposes.
The fee for a HOMER PRO license varies based on the license type (Student, Academic or Standard), billing plan (monthly, annual or other) and required modules. At the moment of research (June 2024) the prices for one month's license with all modules and annual packages vary from about 9 US\$ for students to 380 US\$ for the standard license.

\subsection{Training Effort}\label{Training effort}
The training required for an individual to effectively use a specific software tool is influenced not only by the software tool itself and the available materials to learn it but also by the user's prior experience in modeling energy systems and programming, especially in Python. For users with significant experience in these areas, the training time needed to start using REMix, OEMOF and HOMER PRO is relatively short. Given that REMix and OEMOF are Python-based, users lacking programming proficiency will need more time to model their first energy system using these tools. HOMER PRO, in contrast, doesn't have this disadvantage due to its user interface, making it stand out as the most user-friendly among the four tools studied. According to \cite{CONNOLLY20101059} a typical analysis in HOMER PRO can be completed after one day of training. From our experience, we can support this statement. Furthermore, we estimate that an experienced user could complete their first analysis using OEMOF or REMix by the end of their first working day as well. The training effort for COMANDO is the largest, even if the user is familiar with Python. This is due to a consequence of the high flexibility provided by this tool, which is that the user has to do a lot more coding himself. We estimate, that an experienced user would need two working days to run a typical analysis in this software tool.

In terms of training materials, HOMER PRO is the best equipped. The developers provide more than 20 examples and there is the possibility to purchase some training courses that are offered directly from the developers. For OEMOF there is a large range of teaching content online as well. The number of available tutorials is even larger than HOMER PRO's (about 35). 
The developers of REMix offer 5 tutorials and documentation with fleshed-out explanations to ease the learning curve. Also for COMANDO, there are 5 example projects available. There the developers show special models of energy systems and a documentation of their features. These models further include complex aspects like neural networks, clustering and custom algorithms. The complexity and the amount of explanations inside the Python files show, that the examples are clearly not intended as a tutorial for beginners.

\subsection{Modeling Effort}\label{Modeling Effort}
After learning how to use the software tool, the time required to construct a model with it depends on several factors beyond the user-friendliness of the software. These factors include the complexity of the energy system to be modeled and the specific requirements of the task (such as the precision of the results). The user's level of knowledge about the system they need to model also significantly affects the modeling effort. If the user lacks the necessary knowledge to provide the software tool's inputs, they need to gather more information about the system in order to complete their model. This additional information gathering increases the time and effort required for modeling. The required knowledge level includes, for example, generic information about a module, such as the installation price of a photovoltaic generator. Software tools that can reliably provide such information are a great advantage for users, who do not dispose of them, as they no longer need to search for it to complete the first version of their model.

Among the software tools examined, HOMER PRO is particularly noteworthy in this regard. It offers the most comprehensive amount of supporting information. One example is the capability of accessing solar, wind and temperature data for a given location online, directly from NASA Prediction of Worldwide Energy Resources. 

Such a similar functionality could not be observed in any other of the studied software. Users of OEMOF, COMANDO and REMix have to know or search for usable values for every variable of their energy system model.

Regarding user interface usability, once again HOMER PRO stands out so that it can overall be seen as the software tool with the lowest modeling effort. The requirement to write code in the other tools results in a longer time to design a model and to display or plot the analysis results for these. Due to its high modeling flexibility, the effort to create a model in COMANDO is the highest. There the user needs to create the whole model in Python, define different variable types and formulate the problem to solve on his own.  Consequently, COMANDO requires the user to write more lines of code. In contrast, REMix and OEMOF offer predefined building blocks (such as source, sink, converter and storage), which reduce the programming burden. The user just needs to add parameters and incorporate the components into the energy system with connections. Thus we consider the modeling effort for REMix and OEMOF to be similar, lower than for COMANDO and higher than for HOMER PRO\footnote{As we did not implement any model in any software tool ourselves, we cannot provide a justified quantification of the modeling effort. The scope differences of the software tools further increase the difficulty to find an energy system model, that could be implemented in all four software tools to enable a comparison basis.}.

\subsection{Modeling Scope} \label{modeling scope}

This criterion outlines the types of energy systems that can be modeled with certain software. Key factors that determine the capability of a software tool to model a specific energy system include the supported forms of energy transmission, the range of available components, and the maximum number of components supported for a model. These factors are considered separately in the following paragraphs.

\subsubsection{Forms of energy transmission supported}
From the examples provided, we could notice that REMix supports electrical energy (HVDC or not defined) and gas (CO2 and CH4). COMANDO supports electrical energy (type not defined), heat flows (as a standard household heating system or as a heat flow via mass and enthalpy or heat capacity and temperature) and gas (not defined). From the tutorials analyzed, we could observe that OEMOF supports electrical energy (type is not defined) heat flows, hydrogen, gas (not defined), and different types of other fuels (oil, uranium, biomass, lignite, coal and even mixed fuels). It is noteworthy, that in these three Python software tools, the user can define/name the flow himself so that more flow types than what the examples show are possible. HOMER PRO supports electrical energy (AC and DC, with conversion between those), heat and hydrogen (noticeably, only in HOMER PRO an example with hydrogen storage could be observed). 

\subsubsection{Range of components available}  
Regarding the range of components of each software tool, we analyzed the amount of available commodities and component types, that can be created within a software package. COMANDO stands out in this regard, as it can create the most types of components since the user must design each component and commodity himself. The examples of COMANDO include components like air mass, wall mass, combustion engine as well as compression and absorption chiller. 
REMix and OEMOF provide generic templates, such as storage, source and sink, that can be further defined by the user. For example, electric vehicles can be modeled as battery storage in REMix, or fireplaces can be modeled as wood-to-heat converters in OEMOF. However, more exotic components are difficult to model without making compromises or modifying the modeling system in the system files of the software package. 
HOMER PRO provides a substantial number of predefined components and commodities, including hydroelectric, hydrokinetic, and components for hydrogen systems.  However, it lacks the ability to create custom components and buses like COMANDO or the ability to modify given component templates to the extent possible in OEMOF or REMix.
  
\subsubsection{Maximal number of components within a Model}
Regarding the maximal number of components in a model, for COMANDO, OEMOF and REMix, there is no known limit. The examples of COMANDO and OEMOF have shown what they at least can handle: COMANDO can support at least 15 and OEMOF at least 23. From the documentation of  REMix \cite{REMixsoftwareDocu}, we know that it can support more than a hundred components in its models. 
In contrast, HOMER PRO imposes specific limits depending on the component type. Most components, such as electrolyzer, grid and hydrokinetic can appear on each model only once, normal electric loads can appear twice, batteries, photovoltaic generators and wind generators may come up to 10 times and fuel generators may appear up to 20 times. The biggest model in HOMER PRO could have in total 56 Components. 
One noteworthy consequence of this analysis is that HOMER PRO lacks functionality that the other software packages don't: the ability to model different energy systems separately and aggregate them into one single model. This could be useful for optimization tasks with high detail requirements, such as modeling a neighborhood, where every building disposes of a different photovoltaic, batteries, load profiles, etc.

\subsection{Optimization Goal}
In the analyzed tutorials, we observed differences regarding their optimization goals - that is, the characteristic of the energy system that is to be maximized or minimized in the optimization process. We have categorized the identified optimization goals into four groups, namely economic (minimization of costs), environmental (minimization of impact on the environment), dispatch (minimization of energy losses) and maximization of net power. Table \ref{Optimization Goal Table} provides an overview of the occurrence and frequency of optimization goal categories across the software packages studied.

In REMix the problem is usually formulated as a minimization of costs (especially investment costs), under the condition that the energy demand is met. Therefore the primary optimization goal of this tool can be considered economical. In all five tutorials, a cost for carbon emissions is set, so that also environmental goals can be considered with this software.

In COMANDO the user is required to explicitly create the problem and therefore define the corresponding goal himself. Two of the five examples provided by COMANDO consider economic goals alone and one other considers it together with environmental goals. They mainly include economic goals by optimizing the heating schedule, maximizing the power generation of a specific component or optimizing the investment in an industry district. The remaining 2 tutorials focus on maximizing the net power of an organic Rankine cycle.

The examples provided by the developers of OEMOF demonstrate that this software tool is capable of addressing all four categories of optimization goals. Dispatch optimization is the most recurrent one within 13 of 20 examples. 

HOMER PRO primarily focuses on economic optimization. Similarly to REMix, environmental aspects can also be considered if the user sets a price for carbon emission. This could be observed in one of the twenty-one tutorials available.

\begin{table}
\renewcommand{\arraystretch}{1.3}   
\begin{tabular}{ m{7cm} | *{4}{c}} 

\toprule
        
     Optimization Goal  & REMix & COMANDO & OEMOF & HOMER PRO  \\

\midrule

Economic alone &     - & 2 & 5 & 20 \\
Dispatch &     - & - & 13 & - \\
Environmental alone &     - & - & 1 & - \\
Environmental together with economic  &  5 & 1 & - & 1 \\
Maximize net power  &     - & 2 & 1 & - \\
\midrule
Total examples
&   5 & 5 & 20 & 21 \\
\bottomrule

\end{tabular}
\vspace{3pt}  
\caption{Overview of the presence and frequency of optimization goals categories across the software packages analyzed}
\label{Optimization Goal Table}
\end{table}

\subsection{Optimization Object}
We define optimization objects as the aspects of the model that the software tool can modify or adjust during the optimization process. In our analysis, we identified two main optimization objects that differ across the four software programs studied, namely synthesis and design (see Section \ref{Motivation}) and another less common optimization object: investment path. In this section, we compare the four researched software tools concerning how easily the user could optimize these three objects with their help. 

In all of the software tools studied, design and synthesis optimization can be achieved through simulation and comparison of different variants of the energy system. The tool with which the user can achieve this the easiest is HOMER PRO, as it automatically simulates all possible variations of the energy model given and provides the result of each feasible variant to the user. This way the user can compare each feasible variant of his energy system model relatively easily. Additionally, with HOMER PRO it is possible to set a range of possible values for some variables (the so-called sensitivity analysis) so that even more variants of the energy system can be simulated at once. This helps the user with the design optimization of their energy system. Although the other three software tools also allow the simulation of multiple model variants, the number of variants is considerably small when compared to HOMER PRO.

Another approach for design optimization would be to let the software tool choose the value of a variable (such as the size or number of photovoltaic generators) so that the system is optimal according to the optimization goal. All of the researched software programs are capable of this, whereas, with our analysis approach, no significant difference in the quality or implementation difficulty of this feature could be observed.

Concerning the investment path optimization, this feature was only observed in one of the examples from REMix. With investment path optimization, the user can calculate the optimal investment time points over decades with perfect foresight, if the necessary data is available. No evidence that one of the other three software is capable of a similar feature could be found.

Overall, HOMER PRO stands out in design and synthesis optimization, as it automatically calculates dozens of variants of an energy system and displays the simulation results to the user. The other three software packages are comparable in this aspect, as they offer similar capabilities. Regarding investment path optimization, REMix stands out, as this functionality could not be observed in any other tool investigated.

\section{Discussion} \label{Discussion}
In this Section, we first present the limitations of our approach in Subsection  \ref{Discussion-Approach} that have implications for the results obtained in this work. In the second Subsection \ref{Results}, we analyze the applicability of our results in real world scenarios. In the third and last Subsection \ref{Critical Discussion}, we clarify how the limiting factors of our approach, which are presented in the first subsection, impact the results obtained.

\subsection{Approach} \label{Discussion-Approach}
Our approach allows an initial evaluation of the strengths and weaknesses of the researched software tools and their most suitable application scenario. However, several limitations affect our results:

\begin{itemize}
  \item the limited number of examples provided by the developers and their scope;
  \item the impossibility of running some of the analyzed tutorials (all from REMix and 4 of COMANDO \footnote{We were unable to run the REMix models due to their dependency on GAMS, a licensed software, as explained in Section \ref{Financial Costs}. Regarding COMANDO, we encountered difficulties in correctly installing the solvers needed for these four tutorials, partly because some of them require a license});
  \item the limited workload for this project (about 180 hours); and
  \item the limited information we could find from our literature research, in freely accessible academic documents.
\end{itemize}

The limitations of the approach influence the results obtained in this project, which will be further discussed in Subsection \ref{Critical Discussion}.

\subsection{Results} \label{Results}
Despite the limitations presented in Subsection \ref{Discussion-Approach}, our results can be beneficial to a person in charge of a project to determine if one of these four software tools is suitable for their needs. Moreover, researchers can also make use of the information gathered in this paper to, among other utilities, form a basis or find ideas for further works involving the comparison of software tools.

\begin{table}[htbp]
\centering
\small
\renewcommand{\arraystretch}{1.25}
\setlength{\tabcolsep}{3pt}

\begin{threeparttable}

\begin{tabular}{
    >{\raggedright\arraybackslash}p{1.25cm} |
    >{\centering\arraybackslash}p{2.30cm}
    >{\centering\arraybackslash}p{1.15cm}
    >{\centering\arraybackslash}p{1.15cm}
    >{\centering\arraybackslash}p{1.65cm}
    >{\centering\arraybackslash}p{1.95cm}
    >{\centering\arraybackslash}p{1.25cm}
    >{\centering\arraybackslash}p{2.15cm}
    >{\centering\arraybackslash}p{1.65cm}
}
\toprule

Software
&
Financial costs
&
Training
&
Modeling
&
\multicolumn{3}{c}{Modeling Scope}
&
Optimization
&
Optimization
\\

\cmidrule(lr){5-7}

&
&
Effort
&
Effort
&
\makecell{Energy\\Trans-\\mission}
&
\makecell{Range of\\components}
&
\makecell{Model\\Size}
&
Goal
&
Object
\\

\midrule

REMix
&
\makecell{Open Source,\\solver is\\licensed (\$3500)}
&
1 day\tnote{*}
&
Mid
&
User defined
&
User defined
&
\makecell{User\\defined}
&
\makecell{Economical,\\environmental}
&
\makecell{Design,\\Investment\\Path}
\\

COMAN-DO
&
\makecell{Open Source,\\free solver\\possible}
&
2 days\tnote{*}
&
High
&
User defined
&
\makecell{User-defined,\\highest\\flexibility}
&
\makecell{User\\defined}
&
\makecell{Economical,\\environmental,\\max net power}
&
Design
\\

OEMOF
&
\makecell{Open Source,\\free solver\\possible}
&
1 day\tnote{*}
&
Mid
&
User defined
&
User-defined
&
\makecell{User\\defined}
&
\makecell{Economical,\\environmental,\\dispatch, max\\net power}
&
Design
\\

HOMER PRO
&
\makecell{License\\between \$9 and\\\$380 per month}
&
1 day
&
\makecell{Low\\(GUI)}
&
\makecell{Electric,\\heat, and\\hydrogen\\(predefined)}
&
Predefined
&
Max 56
&
\makecell{Economical,\\environmental}
&
\makecell{Design,\\Synthesis}
\\

\bottomrule
\end{tabular}

\vspace{4pt}

\caption{Overview of the comparison results. Refer to
Section~\ref{Presentation of the Results from the Comparison of the Existing Models} for detailed information}

\label{Overview results}

\begin{tablenotes}[flushleft]
\footnotesize
\item[*] Only if the required programming experience in Python is fulfilled.
\end{tablenotes}

\end{threeparttable}
\end{table}

\subsection{Critical Discussion}\label{Critical Discussion}
In this Subsection, we critically assess how the limitations of the approach outlined in Subsection \ref{Discussion-Approach} impact the results obtained. 

The consequence of a limited number of tutorials and the fact that we were unable to run all of them is that we possibly missed further functionalities of the software tools. Our source of information is limited to the tutorials studied and the documentation consulted. Thus, the software tools are likely capable of more than we could perceive.

The next limitation is the literature and documentation of the software tools, which also impacted the quality of the results: the more relevant information we could find within the literature, the less work we had to conduct ourselves to get the knowledge level we have after the project. This could have led to an expansion of our knowledge beyond our current level shown in this paper due to an adjustment of our workload schedule.

Finally, with a less limited workload, we could have gathered more relevant information and conducted more work. Suggestions regarding the gathering of more relevant information are presented in Subsection \ref{Outlook}.

\section{Conclusion} \label{Conclusion}
In this project, we began by conducting literature research to gather preliminary information on the software tools. Here we also decided to exclude one of the five pre-selected software tools (MTRESS), because it is still in development and we had difficulties finding adequate information about it. Then we made a broad analysis of the example models from all software tools while considering the perspective of a user searching for a suitable software tool to optimize their energy system. From this analysis, some comparison criteria were deduced and subsequently used to compare and evaluate the software tools. A summary of our findings is presented in Table \ref{Overview results}.

These results have then been analyzed to determine the optimal scenario for each of the four remaining software packages. For REMix, this scenario involves an optimization problem with flexible investment paths, where the best path is to be identified. COMANDO is most suitable for energy systems, that require modeling unusual components, such as an air mass or a wall mass. OEMOF should be used for typical energy systems, where minimizing the costs and effort of the optimization task is a priority. Lastly, HOMER PRO is the best choice for projects requiring synthesis optimization of the energy system and also for situations where the responsible for operating the software tool needs to quickly gain insights for further project development and lacks programming experience. Table \ref{table analysis} offers an overview of the analysis of the results.

With this work, we provide information that could help an organization in charge of the optimization of an energy system to evaluate how suitable for their project the studied software packages are. This way we hope to contribute to the optimization of energy systems, as such tasks are crucial for the further development of our society with minimal destruction of the environment and minimal depletion of our natural resources.

\subsection{Outlook} \label{Outlook}
In the future, several possible extensions of our work could be possible to further compare the selected software packages. One potential direction is to identify additional application examples for each software tool to expand our input for comparison under the specified comparison criteria. Furthermore, the comparison criteria could be expanded to include new characteristics of a software package, such as whether and how well the following aspects are considered in their calculations:

\begin{itemize}
  \item maintenance costs,
  \item reliability and repair costs,
  \item performance degradation over time,
  \item escalation of operational costs over time, and
  \item energy losses during transmission.
\end{itemize}

Other characteristics of the software tools that could be studied using new comparison criteria include:
\begin{itemize}
  \item How fast each Framework can perform their calculations/ run their models, and
  \item the minimum and maximum model lifetime supported by each software tool. 
 \end{itemize}

A future step to further enhance the comparison of the software tools studied in this project would be to define an optimization problem and attempt to solve it using different software packages. This approach would allow for a direct comparison of the tools based on their performance in solving a common model or optimization problem. The challenge in this method would be identifying a model or optimization problem that fits within the scopes of multiple software tools simultaneously.

\bibliographystyle{unsrtnat}
\bibliography{main}  

@article{GILS2017173,
title = {Integrated modelling of variable renewable energy-based power supply in Europe},
journal = {Energy},
volume = {123},
pages = {173-188},
year = {2017},
issn = {0360-5442},
doi = {https://doi.org/10.1016/j.energy.2017.01.115},
url = {https://www.sciencedirect.com/science/article/pii/S0360544217301238},
author = {Hans Christian Gils and Yvonne Scholz and Thomas Pregger and Diego {Luca de Tena} and Dominik Heide}
}

@Book{OptimizationEngeneeringBook,
  author =       "Ramteen Sioshansi, Antonio J. Conejo",
  year =         "2017",
  title =        "Optimization in Engineering",
  series =       "Springer Optimization and Its Applications",
  volume =       "",
  edition =      "1",
  address =      "",
  publisher =    "Springer Cham",
  doi =          "https://doi.org/10.1007/978-3-319-56769-3",
  url =          "https://link.springer.com/book/10.1007/978-3-319-56769-3",
  editor =       "",
  number =       "",
  month =        "",
  note =         "(book)",
}

@article{articleFrangopoulos,
author = {Frangopoulos, Christos and von Spakovsky, Michael and Enrico, Sciubba},
year = {2002},
month = {12},
pages = {},
title = {A Brief Review of Methods for the Design and Synthesis Optimization of Energy Systems},
volume = {5},
journal = {International Journal of Thermodynamics},
doi = {10.5541/ijot.97}
}

@article{CONNOLLY20101059,
title = {A review of computer tools for analysing the integration of renewable energy into various energy systems},
journal = {Applied Energy},
volume = {87},
number = {4},
pages = {1059-1082},
year = {2010},
issn = {0306-2619},
doi = {https://doi.org/10.1016/j.apenergy.2009.09.026},
url = {https://www.sciencedirect.com/science/article/pii/S0306261909004188},
author = {D. Connolly and H. Lund and B.V. Mathiesen and M. Leahy}
}

@article{FRANGOPOULOS20181011,
title = {Recent developments and trends in optimization of energy systems},
journal = {Energy},
volume = {164},
pages = {1011-1020},
year = {2018},
issn = {0360-5442},
doi = {https://doi.org/10.1016/j.energy.2018.08.218},
url = {https://www.sciencedirect.com/science/article/pii/S0360544218317547},
author = {Christos A. Frangopoulos}
}

@misc{REMixSoftware,
title = {REMix: A GAMS-based framework for energy system optimization models.},

url = {https://dlr-ve.gitlab.io/esy/remix/framework/dev/index.html},
license =  {BSD-3-Clause},
author = {Manuel Wetzel and Eugenio Salvador Arellano Ruiz and Francesco Witte and Jens Schmugge and Shima Sasanpour and Madhura Yeligeti and Fabia Miorelli and Jan Buschmann and Karl-Kiên Cao and Niklas Wulff and Hedda Gardian and Alexander Rubbert and Benjamin Fuchs and Yvonne Scholz and Hans Christian Gils },
 
lastaccessed = "June 1, 2024",
type = {software},
version = {0.9.7},
year = {2023} 

}

@misc{REMixSoftwareDocu,
title = {REMix: A GAMS-based framework for energy system optimization models.},

note = {https://dlr-ve.gitlab.io/esy/remix/framework/dev/documentation/index.html},
license =  {BSD-3-Clause},
author = {Manuel Wetzel and Eugenio Salvador Arellano Ruiz and Francesco Witte and Jens Schmugge and Shima Sasanpour and Madhura Yeligeti and Fabia Miorelli and Jan Buschmann and Karl-Kiên Cao and Niklas Wulff and Hedda Gardian and Alexander Rubbert and Benjamin Fuchs and Yvonne Scholz and Hans Christian Gils },
 
lastaccessed = "June 1, 2024",
type = {software},
version = {0.9.7},
year = {2023} 

}

@Book{oemofBook,
  author =       "Janet Nagel",
  year =         "1998",
  title =        "Optimierung von Energieversorgungssystemen",
  series =       "",
  volume =       "",
  edition =      "1",
  address =      "",
  publisher =    "Springer Vieweg Cham",
  doi =          "https://doi.org/10.1007/978-3-031-36355-9",
  url =          "https://link.springer.com/book/10.1007/978-3-031-36355-9",
  editor =       "",
  number =       "",
  month =        "",
  note =         "(book)",
}

@misc{HomerPROWebsite,
  author =       "Homer Software",
  year =         "2024",
  title =        "Homer Pro",
  month =        "june",
  lastaccessed = "June 1, 2024",
  note = "https://homerenergy.com/products/pro/index.html",
}

@misc{ComandoHelmholtz,
  author = {Helmholtz Centre Potsdam German Research Centre for Geosciences},
  year =         "2024",
  month =        "july",
  title =        "COMANDO Helmholtz Website",
  lastaccessed = "July 31, 2024",
  note =          "https://helmholtz.software/software/comando",
}

@misc{MtressHelmholtz,
  author = {Helmholtz Centre Potsdam German Research Centre for Geosciences},
  year =         "2024",
  month =        "july",
  title =        "MTRESS Helmholtz Website",
  lastaccessed = "July 31, 2024",
  note =          "https://helmholtz.software/software/mtress",
}

@misc{mtressSoftware,
  doi = {10.5281/ZENODO.11205762},
  url = {https://zenodo.org/doi/10.5281/zenodo.11205762},
  author = {Schönfeldt, Patrik and Schlüters, Sunke and Upadhaya, Ajay and Oltmanns, Keno},
  title = {Model Template for Residential Energy Supply Systems (MTRESS)},
  publisher = {Zenodo},
  year = {2024},
  copyright = {MIT License}
}

@misc{OemofWebsite,
  author =       {Oemof Association},
  year =         "2024",
  month =        "july",
  title =        "OEMOF Website",
  lastaccessed = "July 31, 2024",
  note =          "https://oemof.org/about-oemof/",
}

@article{LANGIU2021107366,
title = {COMANDO: A Next-Generation Open-Source Framework for Energy Systems Optimization},
journal = {Computers \& Chemical Engineering},
volume = {152},
pages = {107366},
year = {2021},
issn = {0098-1354},
doi = {https://doi.org/10.1016/j.compchemeng.2021.107366},
url = {https://www.sciencedirect.com/science/article/pii/S0098135421001447},
author = {Marco Langiu and David Yang Shu and Florian Joseph Baader and Dominik Hering and Uwe Bau and André Xhonneux and Dirk Müller and André Bardow and Alexander Mitsos and Manuel Dahmen}
}

@misc{schönfeldt2022mtress30modell,
      title={MTRESS 3.0 -- Modell Template for Residential Energy Supply Systems}, 
      author={Patrik Schönfeldt and Sunke Schlüters and Keno Oltmanns},
      year={2022},
      eprint={2211.14080},
      archivePrefix={arXiv},
      primaryClass={math.OC},
      url={https://arxiv.org/abs/2211.14080}, 
}

@inbook{PrefaceDeGruyter,
url = {https://doi.org/10.1515/9783110777567-202},
title = {Preface},
booktitle = {Innovations and challenges of the energy transition in smart city districts},
editor = {Sven Leonhardt and Tobias Nusser and Jürgen Görres and Sven Rosinger and Gerhard Stryi-Hipp and Martin Eckhard},
publisher = {De Gruyter},
address = {Berlin, Boston},
pages = {V--VI},
doi = {doi:10.1515/9783110777567-202},
isbn = {9783110777567},
year = {2024},
lastchecked = {2024-07-25}
}






\end{document}